# STRATOSPHERIC BALLOON PAYLOADS FOR ASTRONOMY: THE CHALLENGE OF COPING WITH RISING COMPLEXITY

Vincent Picouet [1,2]

[1] Cahill Center for Astrophysics, California Institute of Technology, Pasadena, CA 91125, USA
[2] Laboratoire d'Astrophysique de Marseille, 38 Rue Frédéric Joliot Curie, F-13013 Marseille, France

ABSTRACT
Stratospheric balloons offer cost-effective access to space and grant the opportunity for fast scientific innovation cycles and higher-risk explorations. In addition to science pathfinders, they serve as platforms for technology advancement, and offer a unique opportunity to train future instrument scientists and PIs. However, the increase in complexity of some projects (sub-arcsecond pointing, numerous degrees of freedom, advanced cooling systems, real-time communication and data transfer, low readiness level technologies, etc.) elevates the scale of challenges. This paper discusses the challenges brought by the increase of instrument complexity in the constrained area of ballooning projects. We use the example of the multi-object slit UV spectrograph FIREBall-2 but also discuss other ambitious payloads to expose some lessons learned. We will then propose strategies for future balloon projects and potential interesting adaptations for funding agencies to accommodate these emerging complexities. The objective being to foster the added value of more ambitious balloon-borne projects, and to initiate discussion about how to generate more robust strategies for handling high-complexity undertakings. Although this paper is enriched by answers from a survey sent to balloon-borne payload PIs, it remains influenced by my personal experience on FIREBall-2. As such, it does not necessarily represent the perspectives of other members of the project, let alone the broader balloon community.

## 1. Introduction

Access to space is essential for observing the Universe in different bandpasses that our atmosphere absorbs (Fig. 1 and 2). Suborbital platforms, such as stratospheric balloons and sounding rockets, offer unique opportunities for scientific exploration due to their cost-effectiveness and flexibility. While sounding rockets cater to relatively small and light niche instruments requiring observations above the atmosphere, balloons can accommodate explorer-sized payloads. This paper delves into the complexities arising from the increasing ambition of balloon-borne astronomy payloads, highlighting their significance, challenges, and potential adaptations to foster their success. We begin by discussing the allure of balloon-borne astronomy and the evolution towards more ambitious projects, followed by an exploration of the rising complexity in payload design and its consequences within the current operational framework. Finally, we suggest adaptations to mitigate some challenges and enhance project success rate.

## 2. Balloon interest

Balloon-borne instruments offer low-cost access to near-space environment. Balloon flights can rival high-profile space missions in several specific domains at a fraction of their cost [1]. The disparity in costs, as illustrated in Tab. 1, is predominantly attributed to launch expenses. Unlike satellite missions, which entail significant infrastructure investments and higher risk aversion due to their important costs, balloon and rocket projects often leverage the expertise of students and postdocs, thus minimizing labor expenses. Furthermore, the relatively smaller scale of balloon and rocket projects means they almost never necessitate the construction of dedicated space infrastructure at the launch site, further contributing to their cost-effectiveness.

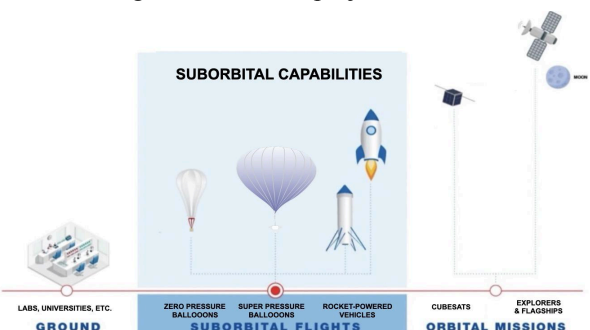

Figure 1: Schematic diagram of astronomical observing capabilities [Adapted from NASA].

| Criterion | Ground based | Sounding Rockets | Zero pressure Balloons | Super pressure Balloons | SmallSats | Explorers | Flagships |
|---|---|---|---|---|---|---|---|
| Bandpass interest | VIS - IR Micro - Radio | X-ray, UV, IR | UV, IR, Microwave, Gamma & X ray | | All | | |
| Cost | Cheaper | ~$5M | ~$10M* | ~$10M* | ~$5-20M | ~$100-250M | >$10B |
| Average flight time | N/A | 5-10 minutes | 3h → 2 days | up to 100 days | ~1 year | 2+ years | 5+ years |
| Altitude | <5km | 100-300 km | 30-40 km | 20-35 km | Space | | |
| Max weight | | 500 kg | 2000 kg | 1000 kg | 5-50 kg | 300-1500 kg | 5000 kg |
| Max size | Variable | ~5×0.5 m² | ~30 m³ | | 3e-2 m³ | ~30m³ | ~100 m³ |
| Project timeline | | 2-5 yrs | 2-5 yrs* | | 1-5 yrs | 7-10 yrs | 10-15 yrs |
| Frequency | Variable | ~10 yr⁻¹ | ~10 yr⁻¹ | 2 yr⁻¹ | 1 yr⁻¹ | Every few yrs | 1/decade |
| Risk aversion | Low | Low | Low* | Medium* | Low/Medium | Medium/High | High |

Table 1: Specificities of the different astronomical observing capabilities. Colors show strength and limitations of each observing capability.
* Specificities impacted by projects rising complexity

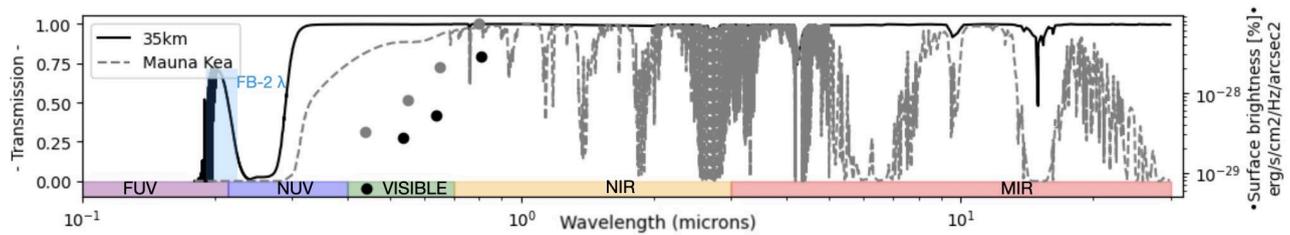

*Figure 2: Atmospheric transmission and surface brightness (SB) at different operational altitudes [2, 3]*

While in France technology advancement and training are mostly seen as byproducts behind real science drivers, NASA evaluates similarly suborbital proposals based on Science, Technology & Training. This can lead to significant differences between the French and US suborbital programs.

**2.1.** Technology readiness level advancement and proof of concept

The affordability and minimal resource requirements of these projects offer numerous advantages. They facilitate rapid iteration cycles, with many payloads capable of being launched every one or two years. Consequently, these projects are more willing to accept higher levels of risk, fostering incremental iterative improvements over time. This approach is very valuable for advancing the readiness level of critical technologies and validating instrumental concepts. For instance, it allows for the testing of technologies before their integration into satellite missions, thereby mitigating potential risks. Additionally, it provides opportunities to demonstrate the feasibility and effectiveness of various instrument designs and functionalities, paving the way for their eventual deployment in space-based missions. Astrophysics explorers since 2000 offer a compelling example, with a significant majority of these missions having benefited from instrumental heritage originating in suborbital projects [4].

**2.2.** Training and personnel resource accelerator

The rapid development timelines associated with balloon projects present a unique opportunity to train the next generation of scientists and engineers. These projects serve as an invaluable training ground for students and young professionals, offering hands-on experience in mission design, project management, and cutting-edge scientific research.
While the inclusion of student piggy-back payloads on program flights provides college students with practical training opportunities [5], engagement in bigger balloon projects also allows graduate students and young scientists to immerse themselves in all phases of payload development, from initial scientific concept formulation to mission design, construction, integration, testing, flight operations, and data analysis. This comprehensive experience not only enriches their academic pursuits but also prepares them for future roles in the field of space science and engineering [5]. Notably, the Principal Investigators of all astrophysics explorers since 2000 (including UVEX, COSI, SPHEREX, IXPE, TESS, NuSTAR, WISE, Swift, and GALEX) have had some heritage in suborbital instruments [4].

**2.3.** Science

While sounding rockets and CubeSats also offer valuable platforms for experimentation, balloon payloads provide distinct advantages, and are, in particular, the only vehicles allowing explorer mass/size payloads allowing large collecting area during an extended observation time. Therefore, balloons enable scientists to achieve higher scientific yield relative to the effort invested, making them highly appealing for a wide range of astronomy fields.
Having a 2 tons explorer size instrument at 35 km altitude for a few nights is of huge interest for several key sciences such as UV and infrared astronomy, CMB science, High-Energy Astrophysics, Particle Astrophysics, Solar Physics, Geospace Science, Earth Science. The science enabling possibilities they allow have already been shown many times with for instance the discovery in the 1970s of gamma-ray line emission at 511 keV from the annihilation of positrons in the galactic interstellar medium or the BOOMERanG (1.3m primary, 0.3K cooled bolometric detectors) and MAXIMA instruments data [6] which provided the first undisputed evidence that the Universe has a flat geometry, before PLANCK.
Currently, NASA and CNES respectively fly ~12 and ~6 large payloads on conventional (~2 hours to 2 days) balloon flights per year with relatively low failure rates (Fig. 3) and a few long-duration balloons (3-41 days) each year. They now have the capability of using super-long-duration balloons allowing flights to up to 60 days. This introduction of super pressure Ultra-Long Duration Balloons (ULDB) represents a significant advancement that should finally allow balloon payloads to reach scientific goals historically only accessible by satellites [1].

**2.4.** The interest of more sophisticated / complex / ambitious payloads

The evolution of balloon-borne missions from relatively simple laboratory experiments in the 60s to the apparition of more sophisticated payloads reflects a significant shift in scientific ballooning, driven mostly by the advancement of ballooning technology, and in particular by three interconnected aspects: the ability to carry **larger weight**, the progressive **decline of balloon failures rate** (Fig. 3) and the ability to perform **longer flights**. This both increased the scientific appeal of balloon platforms and allowed the development of more sophisticated experiments.

The current science enabling capability of balloons platforms, led several astronomy-related projects to elevate the complexity of their project to achieve more challenging astronomy goals (eg. MAXIMA, FIREBall-2, GUSTO), requiring higher performance (resolution, pointing, field of view, surface brightness limit, cryostat), more complex optical system with higher degrees of freedom.

The adoption of complex balloon payloads not only augments scientific objectives but also plays a pivotal role in training scientists to the more and more complex astronomical instruments (an objective highlighted by NASA & the NAS NRC [6] [5]). With the constant sophistication of astronomy orbital projects always taking up more complex scientific and technological challenges, developing more sophisticated suborbital payloads can be really beneficial. The development of ultra-long-duration balloons even accentuates the potential interest of sophisticated payloads.

More ambitious payloads, combined with the flexibility of suborbital projects could have a real positive impact. In addition to the example of ambitious stratospheric project which became precursor to Satellite Missions, the exigencies of complex balloon projects compel teams to develop streamlined approaches to problem-solving, fostering innovation and efficiency within the stratospheric research domain.

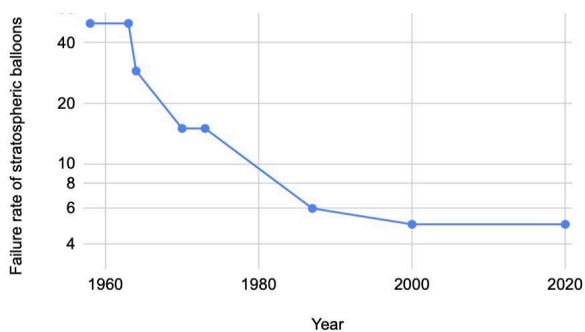

*Figure 3: Progressive decline of balloon failure rate at CSBF, currently plateauing at 5% in average [7], most likely higher for the largest zero pressure balloons and the super pressure ULDB ones.*

3. Development of sophisticated astronomy payloads

For the reasons cited above, there is a continual emergence of more ambitious playloads requiring sophisticated subsystems, technology and integrations:
- Dedicated ≥2-axis sub-arcsecond pointing
- Larger collecting area ➤ increasing weight & cost
- Complex optical systems
- Dedicated and sophisticated cooling system
- High number of degrees of freedom
- New subsystems required for long duration flights (efficient solar panels, batteries & cooling systems), increasing complexity and payload weight
- Dedicated downlink/telemetry capability
- More complex and longer integration

This is in addition to the already complex stratospheric environment: altitude winds, significant pendulum required to be accurately compensated for, highly changing environment ($T_{ext} \in [-70, +40]$°C, $P_{ext} \in [0.003, 1000]$mBar) requiring efficient thermal control and generating mechanical deformation, strong deceleration after parachute opening and rough impact at landing (~10g), etc.

Consequently, the alignment of ambitious project objectives with the existing constraints (lower funding and human resource) and organizational frameworks (less system engineering, high turnover) of suborbital missions can present a significant hurdle. While this demanding framework actually yields significant innovations to overcome these hard constraints, complex instruments require more development time and the risks associated with instrument complexity can increase rapidly. If these projects continue to advance TRL efficiently, proof of concept of more complex instruments and scientific returns can be more challenging to reach in the existing current framework of stratospheric balloons.

**3.1.** FIREBall-2 example

The Faint Intergalactic Redshifted Emission Balloon (FIREBall-2, FB-2) is a one-meter telescope coupled to a UV multi-object spectrograph (MOS) designed to image the low-redshift surrounding gas of galaxies in emission at low cost compared to orbital projects [8]. The instrument presents an interesting case study of the increase of complexity in a balloon-borne experiment. The instrument, that can still be comprehended and optimized as a whole, is at the limit of the complexity of what is achievable in the limited budget of suborbital instruments.

FIREBall-2 performed the very first multi-slit acquisition from space - any band, and besides the difficulty to reach strong scientific results, the project holds interesting lessons for future balloon-borne missions. The following sections will discuss both some of the instrument's complexities as well as the additional complexity associated with designing it in the current suborbital organizational frameworks.

### 3.1.1. Design

We list in Tab. 2 some of the many complex aspects of the FIREBall-2 instrument.

| Low surface brightness instrument | Multi 70-μm slits spectrograph |
|---|---|
| • Background must stay below a few e⁻/pix//h<br>• Requires extremely careful baffling<br>• Large collecting area & high-altitude long observing night (+ moon constraints) | • Requires sub-arcsecond guidance<br>• Extremely demanding XY calibration<br>• ~1m³ instrument rotation in gondola<br>• Many degrees of freedom |
| **Complex optical system (10 optics)** | **Cooled low-TRL detector** |
| • 2 × 1mØ mirrors - degraded at landing, challenging to characterize for flight<br>• Whole 5m gondola is part of optical bench<br>• Fast telescope ➔ Controlled mech. deformation | • 10⁻⁶ mb pressurized instrument<br>• Intricate detector requiring fine tuning/process.<br>• Designed end-to-end cooling system capable of dissipating 35MJ |
| **Launch & flight operations** | **Challenging calibration at the launch base** |
| • Altitude & duration req.+ Moon constraints<br>• Long acquisition procedures | • 1T of optics/subsystem to AIT<br>• UV ➔ Challenging testing on earth |

*Table 2: List of FIREBall instrument's complexities*

In addition to these difficulties, the instrument presents an important number of degrees of freedom that complicates the instrument design and introduces risks. Most of the DoFs failed or generated issues in one of the flight or integration campaigns:
- Azimuth pivot was damaged during FB-1 first launch
- Rotation stage: failed at the end of the 1$^{st}$ FB-2 flight
- Slit mask wheel: generated ~2e⁻/h straylight issue potentially due to an optical sensor
- Tip-tilt focus actuators below the tank: several failures during integration currently unexplained
- 2-axis siderostat (~0.3" axis RMS control): no issue, likely due to several years development and testing

### 3.1.2. Integration & management

Complex instruments and gondola require longer assembly and integration time as well as extensive fully-integrated testing and calibration at the launch base. In the case of FIREBall-2, some optical subsystems (e.g. focal-corrector assembly) are so demanding in terms of alignment (a few tens of micron) that regular metrology (by metrology arm or non surface contact scanning technology) is not sufficient. It then requires a long and demanding interferometric verification (and potential re-alignment) to reach the subsystem alignment tolerance.

Instruments with a high level of complexity and sophisticated integration and calibration requirements require extensive interfacing and optimization that must be verified/repeated each time the payload is moved to a new location, including at the launch site. This can be extremely challenging given the absence of clean room and very limited ground support equipment (GSE). Fast optical instruments, characterized by a low $f$-ratio, in addition to alignment sensitivity, are less tolerant to drift, instabilities or thermal expansion. On the FIREBall-2 instrument, a change of environmental temperature of one celsius degrees during the final 4-hour long platescale ground calibration would invalidate the results.

### 3.1.3. Management and personpower

The typical organization level of suborbital payloads, which is much less stringent than that of orbital projects and offers essential flexibility, might be insufficient for longer and more complex projects that involve multiple sophisticated subsystems and require extended learning curves. The currently inevitable student or postdoc turnover, is more challenging to handle on ambitious and long-development payloads and requires a higher level of organization. For FIREBall, continuity has been a challenge which can be heavily affected by the departure of experienced key personnel. The project would also have benefitted from having support from an experienced system engineer specialized in stratospheric environment, either as part of the team or provided by the suborbital program for punctual support. This would allow the team to not only learn from failures, but from experts along the way and would also help build heritage and continuity as well as only taking sensible risks. Complex payloads need complex situational awareness to help assess failures, a factor that less experienced scientists may overlook.

### 3.1.4. Launches

**Launch unpredictability:** Unlike sounding rocket launches, which are typically scheduled with precision, payload launches using large stratospheric balloons remain extremely unpredictable and can lead to significant delays which can reduce the attractiveness compared to the effort that goes into designing such payloads. This unpredictability stems from various factors, including weather conditions, wind directions and speed, logistical/operational challenges/security, increasingly rigorous security and trajectory constraints, etc. As an example, FB2 had only one single flight opportunity during the 2023 campaign (2 in 2018). This can rapidly prevent complex instruments (like low SB ones) to respect other flight opportunity constraints which would reduce further or even suppress any suitable launch windows, such as minimum duration, altitude requirements or moon phase/position constraints. The oversubscription of ready-to-launch payloads vs. launch opportunities forces some payloads to leave the base without flying.

**Balloon failures:** Aerostar and the Columbia Scientific Balloon Facility (CSBF) are invaluable partners in allowing stratospheric balloon missions. However, recent experiences have highlighted opportunities for enhancement, particularly for missions with higher complexity and cost. FIREBall flights from Fort Sumner revealed several challenges like launch operational issues (2007 & 2018 flight) or balloon integrity problems, leading to premature terminations (2018, 2023). The first balloon failure, which was due to a systematic issue generating too much stress on the balloons leading to early termination for several payloads including FB-2 [8]. While it was identified and fixed FIREBall-2 still experienced another balloon failure in 2023 from an unknown cause, possibly attributable to the random failure rate of the 39MCF balloons or the unusual wind direction that required to unfold the balloon over a less clean/optimal launch area. These issues can significantly increase projects' failure risks and therefore scientific appeal.

While acknowledging the strong challenging constraints faced by CSBF, such as limited funding, high operational complexities and important turnover [9], it is crucial to explore avenues for improving support mechanisms, including bolstering equipment maintenance (and potential balloon ranking), optimizing launch scheduling, and fostering closer collaboration between CSBF and project teams from launch show-ups to termination. Addressing these concerns collaboratively and defining a clear line of communication between science teams and launch teams would help strengthen the viability of stratospheric balloon missions, which seems essential to support the emergence of more expensive and sophisticated playloads.

### 3.2. Implications

While the interest of more scientifically ambitious balloon-borne payloads is appealing, managing to elevate the complexity of payloads within the current constraints of stratospheric platforms is challenging. For these projects, the number of complex subsystems under the responsibility of the scientific teams is very high (such as the cooling system, part of the guidance and communication systems, solar panels, etc. in addition to the payload subsystems). As a result, these teams may be undersized to maintain a sufficient level of reliability across all subsystems. It is also hard to stay competitive with other less-complex payloads as more ambitious and sophisticated payloads usually weaken the historical interest of stratospheric instruments: they require longer development and repair time between launches, they are more expensive and can have higher levels of risks. In these constraints, experimental groups can struggle to find funding for the detector development, optics design, and cryogenic engineering required to construct scientifically competitive payloads [5].

Whereas the current situation does not appear to be fully adapted for the development and success of more complex payloads, it seems that some relatively simple changes could simplify the emergence of more ambitious projects without reducing their ability to reach their scientific objectives.

### 4. Path forward / adaptations

Drawing from our experience with a highly sophisticated astrophysical payload on which we encountered numerous unforeseen challenges, and using some feedback from other ambitious projects such as PILOT, Archeops, PICTURE-C/D, FOCA, SuperBIT, and Spider, we will explore potential avenues to facilitate the emergence of more ambitious payloads, and increase their chances of success. It is, of course, not possible to make an exhaustive list of considerations for improving the success rate of more complex balloon-borne projects. It would be very multi-dimensional (requiring highly experienced team and PI, extremely thorough testing, etc) and would still, unfortunately, rely to a significant extent on factors beyond one's control, such as launch opportunities, balloon failures, single point failures, damage at landing, etc. Still, in the following subsections we try to emphasize a few pieces of advice to try to increase success rate when transitioning from relatively simple payloads to more sophisticated ones, both for projects and launching/funding agencies.

They can seem obvious with some perspective, as they almost all derive from the fact that for more ambitious projects with complexity between average suborbital payloads and orbital payloads, the different tradeoffs (organization vs. flexibility, training vs. expertise, etc., see Fig. 3) can not be the same as average complexity suborbital payloads. Though, these suggestions can currently be challenging to implement with the current constraints of stratospheric projects organization.

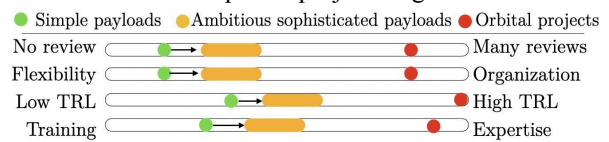

*Figure 3: Diagram illustrating the need of refining risk aversion and resulting tradeoffs for more sophisticated/expensive payloads in order to keep the same level of total integrated risk.*

**4.1.** For projects
    4.1.1. Organization
        4.1.1.1. Person power

In more ambitious suborbital projects, it is particularly important to find the right ratio between training and expertise needed within the project. The same way this optimal ratio is different for orbital and suborbital projects, it must be different for piggyback experiments, relatively low complexity projects and very sophisticated payloads. To maintain a similar success rate for sophisticated payloads than simpler projects, projects can require higher experienced people and specialized engineers. For projects leaning towards employing younger scientists — graduate students and postdocs — for training, the potential loss of expertise can be mitigated by providing more support or oversight from the PI or senior system engineer.

The significant use of graduate students and postdocs can lead to additional obstacles due to the important turnover of a few years. It can be important to try reducing this turnover to improve continuity and build a deeper understanding from sophisticated projects. In any case, it is essential, before facing any turnover or departure, to extensively plan and organize knowledge transfer. Ambitious projects come with longer learning curves. Therefore, any significant person power change, can require several months of overlap for the newcomer to be properly trained. Likewise, having students/scientists spread out on too many projects can reduce the depth of understanding of the instrument.

### 4.1.1.2. Project organization

Similarly, more ambitious projects could benefit from some project organization changes in order to keep a relatively high success rate (Fig. 5).
- Have access to specialized engineers for the most complex subsystems of the instrument
- Limit/minimize the number of low-TRL technologies, high-complexity subsystems, degrees of freedom, single point failures
- Decide knowingly how to handle the tradeoff between flexibility/rapidness and more demanding organization/documentation
- Use more strict space qualification for critical systems (on-board computers, cooling systems, etc.)
- Almost all projects argued that reviews were not helpful to identify problems on the instrument side. Optimize these reviews (and potentially include scientists from Balloon Working Groups) could be valuable to better gauge mission feasibility, verify nothing is overlooked and provide constructive feedback.

Other factors can potentially influence the success of more ambitious stratospheric projects. Iterative improvement of balloons increases chances of success. This might have been particularly true on instruments that could be flown every couple of years, such as the iterative development of SCAP ➛ FOCA ➛ GALEX or BOOMERanG ➛ MAXIMA ➛ PLANCK. In the same vein, projects that are designed as path finders for bigger projects, might benefit from more support, especially if the orbital project is already under development (eg. PLANCK).

Ambitious projects that entail higher costs often involve resources pooled from multiple institutions, and even across different countries. While this collaborative approach can help mitigate expenses for individual institutions or agencies, and provide access to diverse expertise, it also introduces complexities such as management, logistics, integration, and synergy challenges. These factors can elevate the overall project costs and diminish significantly flexibility, and should therefore be carefully considered.

Another basic but important aspect that can potentially compensate for part of the above is the extensive use of modular design. This allows to simplify the project management if several institutions are present. It will also allow easier iterative improvement/repair of the instrument, especially after a flight, will offer better testing opportunities and more flexibility in dealing with unexpected developments. The question of wanting to refly the payload and gondola as well as the minimal required AIT to make it re-fly should be thoroughly thought through during development as it will have a big impact on required time/funding requirements. The design of the instrument should depend on this evaluation and modular design can help keep a higher fraction of invested development effort. Despite these suggestions, one of the biggest strengths of balloon borne astronomy is the flexibility it allows. If more ambitious projects might require more organization to reach a successful flight, it is essential to still try to keep significant flexibility to be able to face the many challenges along the way.

### 4.1.2. Extensive testing and risk management

One of the most essential paths to success for a sophisticated payload is thorough testing and a deep understanding of the risks. More ambitious science goals require a more precise definition of the requirements and therefore a much finer set of tests. Therefore, projects can really beneficiate from support from an experienced systems engineer to help:
- Implementing a comprehensive risk management strategy for the project as well as identifying potential pitfalls. Develop mitigation strategies and design contingency plans
- Defining a clear alignment/integration/test plan, compatible with the tolerancing/requirements and all the others project constraints
- Evaluating accurately instrument performance using stringent and unbiased estimators

On stratospheric payloads, projects could assume that any non-tested subsystem won't work as expected. Each test should be as close as possible as flight conditions. While end-to-end tests are demanding but very useful they do not allow to disentangle all the performance loss, which makes subsystem tests also an important aspect to consider. Slightly more systematic risk/performance assessments would also be beneficial to the projects. Having an easy to use instrument model that allows analyzing the instrument end-to-end performance, to perform the trade studies and optimize the instrument tuning can be extremely helpful.

**4.2.** For funding and launching agencies

### 4.2.1. Support

While it seems consensual that the most scientifically ambitious project will need more funding in order to be developed [5 - 6], there are also funding-independent changes that launching and funding agencies can implement to improve science/technological success of these projects, as well as training, to improve their return on investment.

A clear output from discussions was the interest of having a referent system engineer with deep sub-orbital experience. This point of contact for projects, would allow to increase support, review some of the instrumentation and provide testing requirements to ensure the equipment will function properly during flight. This would also allow agencies to concatenate expertise in stratospheric instrumentation (eg, on-board computers, thermal/power systems, etc.). This is not only important for projects, but also in the long run for agencies to be sure that the knowledge/expertise required to allow incremental improvement is secured. Coupled with more serious reviews, this would allow the detection of potential issues much more rapidly.

### 4.2.2. Hardware and equipment

Some equipment, subsystems and documentation would also greatly benefit from improvements to simplify the development of more ambitious astronomy balloon borne projects:

- The number of successful balloon flights has dropped over the past years [9]. Work on reducing both systematic & random balloon failures seems essential
- Balloon launches are still too unpredictable in US bases compared to the effort that goes into preparing a payload. Developing the possibility of increasing launch rates (especially for 24h+ payloads) would be very beneficial (develop alternative US launch site?)[1]
- Clearer interface document control describing all gondola requirements [1]
- Increase possibility of recovering payloads for SPB instruments and enhance crash-pads capabilities to improve all recovery capabilities.
- Provide complete communication solution with higher telemetry/downlink performance[1]
- Provide and improve stratospheric platforms to reduce risks (sub-arcseconds pointing gondolas, batteries/solar panels cooling system)[1] and reduce workload on scientific teams
- Launch bases would benefit from having a clean room/tent capability as the final end-to-end integration/testing of the payloads happens there.

With the consolidation of stratospheric ballooning and the standardization of the needs, if launching agencies can not support the development of the demand (pointing systems, communication, cooling systems), hopefully the private sector (eg. STARSPEC) could potentially get interested in developing generic and robust solutions for stratospheric platforms.

### 4.3. Expertise aggregation

Another essential aspect is the capability to capitalize on knowledge and expertise. As scientific instruments become progressively more complex to reach more ambitious science goals, scientific teams cannot afford to reinvent the wheel on all aspects without compromising their ability to succeed.

Currently, it appears that the only response provided to this difficult challenge is the organization of suborbital symposiums and the establishment of working groups, often formed by instrument principal investigators (PIs), who, on their free time, can provide inputs to space agencies management on the needs and requirements from the balloon users. While such a working group can provide extremely useful recommendations [10] on the ballooning aspect, its goal is not to improve knowledge capitalization or continuity.

Some effort should also be put on the development and organization of comprehensive feedback from balloon borne projects, improve the archive or balloon borne instrument and launches [11], document the lessons learned, pitfalls, and successful approaches, publish advances on balloon technology and payload instrumentation[2]. This approach, if implemented strategically, should help reduce failure rates, increase return on investment, expand project leads networks, raise balloon-borne instrument scientist awareness about stratospheric constraints and pitfalls, reduce knowledge redundancies, and provide access to point of contact on specific suborbital domains. It also aims to increase opportunities to leverage cross-cutting capabilities and technologies across the balloon-borne instruments portfolio.

While this expertise/knowledge aggregation could be implemented at several levels (from small to large: scientific/project teams, research laboratories, launching agencies, funding agencies), the most sensible way is for the biggest stakeholder to lead and organize this effort.

In the United States, competition for grants between research teams or universities can incentivize them to withhold knowledge to remain competitive and secure their funding, making such initiatives challenging.

Despite the potential detrimental effects both scientifically and economically, funding agencies have not yet taken the role of organizing such efforts. However, a promising initiative for cubesat projects, with the inauguration of the Cubesat Learn Forum in 2022, may lead to positive outcomes and hopefully inspire the generalization of similar concepts to the balloon (and sounding rocket) suborbital programs.

---

[1] These recommendations were also done by the Balloon program analysis group [10]

[2] Astro2020 advocated for more guidance, common hardware/software as well as the development of an outreach program [9]

The absence of strong conflict of interest in the field and the job security in research in France should simplify this task, potentially leaving the role of organizing this effort to larger institutions, such as launching/funding agencies.

## 5. Conclusion

Stratospheric balloons are still mostly used for relatively moderate complex payloads in areas going from astronomy, particle physics, geospace science, earth science. This explains the relatively high funding constraints. However, the potential to compete with space projects in some specific astronomy fields both explains and justify the development of more ambitious balloon-borne projects. This trend is further fueled by advancements such as super-long-duration balloons, enabling extended observation times of up to 100 nights. Recognizing the potential of these endeavors, both NASA [6] and the National Academies of Sciences [5] have advocated for aligning funding to support sophisticated astronomy-related science payloads which might have led to the development of the $20M-cost-cap pioneer program. Based on the experience on a sophisticated payload, this paper has discussed funding-independent pathways to facilitate the development of ambitious and sophisticated payloads, both for the project side and funding/launching agencies. Most of these recommendations are summarized Fig. 5. Because flexibility is a major benefit of suborbital programs, the real difficulty is to find a way of knowing where to position the flexibility/organization tradeoff cursor. Having the capability to estimate balloon borne projects' complexity (by designing a metric for instance) could help implementing an appropriate response.

## 6. Acknowledgments

FIREBall-2 is co-funded by CNES and NASA. Ground support was provided by the CSBF during the different integration/flight campaigns. This work benefited from discussions with Drew Miles, Matt Matuszewski, Erika Hamden, Johan Montel, Bruno Milliard, William Jones (SuperBit PI), Christopher Mendillo (Picture-C/D PI), Alain Benoit (Archeops PI), JP Bernard (PILOT PI), Luis E.Pacheco (Astrocat Website).

## 7. References

1. Cho A. (2020) Cheap balloon-borne telescopes aim to rival space observatories, Science, 367, 967
2. Kremic T. (2015) Stratospheric Balloons for Planetary Science and the Balloon Observation Platform for Planetary Science
3. Gill A., et al. (2020) Optical night sky brightness measurements from the stratosphere, AJ
4. Miles et al. *in prep.* JATIS special issue
5. National Academies of Sciences, Engineering, and Medicine (2010) Revitalizing NASA's Suborbital Program: Advancing Science, Driving Innovation, and Developing Workforce | The National Academies Press.
6. NASA Stratospheric Balloons: Science at the edge of space – Report of the scientific ballooning assessment group (2010)
7. Bawcom D. A short history of the NSBF (CSBF)
8. Picouet V., Milliard B., Kyne G., Vibert D., Schiminovich D., Martin C., Hamden E., et al., (2020) JATIS, 6, 044004.
9. Pathways to Discovery in Astronomy and Astrophysics for the 2020s. National Academies Press. https://doi.org/10.17226/26141, AAS 2021
10. Report of the Balloon Program Analysis Group (2020)
11. Balloon launches database: https://stratocat.com.ar

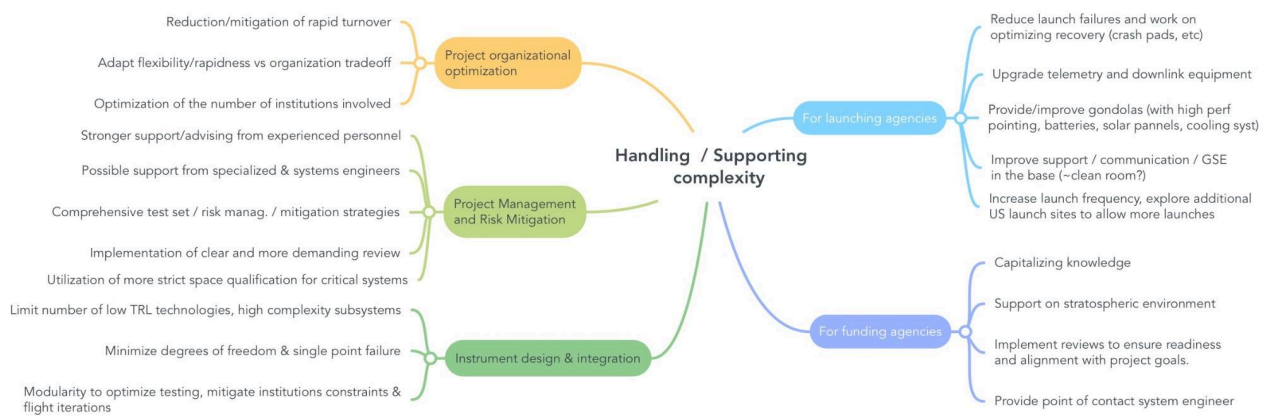

*Figure 5: List of funding-independent changes that could participate to improve success rate and return on investment of more ambitious or sophisticated balloon-borne payloads.*